\documentclass[11pt]{article}
\usepackage{graphicx}
\begin{document}
\vspace*{-.6in}
\thispagestyle{empty}
\begin{flushright}
hep-th/0603197
\end{flushright}
\baselineskip = 18pt
\renewcommand{\baselinestretch}{.88}

\vspace{.5in} {\Large
\begin{center}
\textbf{Mass and Thermodynamics of Kaluza-Klein Black Holes with
Squashed Horizons}
\end{center}}
\vspace*{1cm}

\begin{center}
Rong-Gen Cai,$^{a,c,}$\footnote{Email address: cairg@itp.ac.cn}
Li-Ming Cao$^{a,b,}$\footnote{Email address: caolm@itp.ac.cn} and
Nobuyoshi Ohta$^{c,}$\footnote{Email address:
ohta@phys.sci.osaka-u.ac.jp.
Address after 31 March 2006:
Department of Physics, Kinki University,
Higashi-Osaka, Osaka 577-8502, Japan}
\end{center}

\begin{center}
\emph{$^{a}$ Institute of Theoretical Physics, Chinese Academy
of Sciences, \\ P.O. Box 2735, Beijing 100080, China \\
$^{b}$ Graduate School of the Chinese Academy of Sciences,
Beijing 100039, China \\
$^c$ Department of Physics, Osaka University, Toyonaka, Osaka
560-0043, Japan}

\vspace{.5in}


\end{center}

\vspace{.2in}

\begin{center}
\underline{ABSTRACT}
\end{center}
\begin{quotation}
\noindent Recently a five-dimensional Kaluza-Klein black hole
solution with squashed horizon has been found in hep-th/0510094.
The black hole spacetime is asymptotically locally flat and has a
spatial infinity  $S^1 \hookrightarrow S^{2}$.  By using
``boundary counterterm" method and generalized  Abbott-Deser
method, we calculate the mass of this black hole. When an
appropriate background is chosen, the generalized Abbott-Deser
method gives the same mass as the ``boundary counterterm" method.
The mass is found to satisfy the first law of black hole
thermodynamics. The thermodynamic properties of the Kaluza-Klein
black hole are discussed and are compared to those of its
undeformed counterpart, a five-dimensional Reissner-Nordstr\"om
black hole.

\end{quotation}

\vfil
\centerline{}

\newpage

\addtocontents{toc}{\protect\setcounter{tocdepth}{2}}
\pagenumbering{arabic}

\vspace{.5in}
\section{Introduction}
Looking for exact black hole solutions in the Einstein-Maxwell
theory (with other possible matters) is a subject of long standing
interest. Indeed over the past years we have witnessed a surge of
new black hole solutions found in general relativity and
supergravity theory. In four dimensions, when the dominant
energy condition is satisfied, Hawking~\cite{Haw} has shown that
the horizon of a black hole must be a two-dimensional round sphere
$S^2$. In higher ($D>4$) dimensional spacetimes, the topology of
black hole horizon is rather rich. To give an example in five
dimensions, one can have a black hole with horizon topology $S^3$
and black string with horizon topology $S^2\times R(S^1)$.
Recently Emparan and Reall~\cite{ER} have found a rotating black
ring solution with horizon topology $S^2\times S^1$.

 In the five-dimensional Einstein-Maxwell theory, very recently
 Ishihara and Matsuno  have found a
charged black hole (named Kaluza-Klein black hole) solution which
has a spatial infinity of squashed sphere or $S^{1}$ bundle over
$S^{2}$~\cite{Ishihara}. This solution has a lot of interesting
properties as discussed by the authors. Due to its topology
difference from the usual black holes, it deserves further study.
In this note we will calculate the gravitational mass of the black
hole and then discuss its thermodynamic properties.

To study the thermodynamics of a black hole, one should first give
the conserved charges of the system. Many methods have been
developed to calculate the conserved charges of gravitational
configurations in recent years. But most of those methods are
background dependent, for example, Abott-Deser method~\cite{AD},
Euclidean action method~\cite{GH} and N\"other method. To get a
finite result, one has to select a reference solution which has
the same boundary geometry as the gravitational configuration
under consideration. On the other hand, choosing different
background will lead to different result in some cases. In the
domain of asymptotically AdS space-times or asymptotically locally
AdS space-times, motivated by AdS/CFT correspondence,  a
background independent method, called ``boundary counterterm"
method, has been proposed~\cite{bk, Kraus}.  Furthermore in the
same paper~\cite{Kraus}, Kraus {\it et al.} have also suggested
the counterterms for 4- and 5-dimensional asymptotically flat
space-times with boundary $R^2\times S^2$ or $R\times S^3$ and
given a formula to calculate the conserved charges. In a recent
work~\cite{Mann}, Mann and Stelea have proposed a counterterm in
5-dimensional space-times with boundary topology
$R^2\hookrightarrow S^2$. This counterterm is equivalent to the
counterterm of Kraus {\it et al.} when the space-times have the
boundary topology $R^2\times S^2$. Using this counterterm they
have calculated the mass of the Kaluza-Klein monopole (with a
boundary topology $R\times S^1 \hookrightarrow S^2$), which is in
agreement with the results previously obtained in the literature.
We  find that the Kaluza-Klein black hole found in \cite{Ishihara}
has the same boundary topology as the Kaluza-Klein monopole. Thus
it is natural to use this method to calculate the mass of the
black hole.

This paper is organized as follows.  In the next section, we will
briefly review the Kaluza-Klein black hole with squashed horizon.
In Sec.~3, we calculate the mass of the black hole using the
counterterm method and generalized Abott-Deser method. In Sec.~4,
we study the thermodynamic properties of this black hole and find
this mass defined in our way satisfies the first law of black hole
thermodynamics.  We end in Sec.~5 with our conclusion and
discussion.

\section{ A Brief Review of the Kaluza-Klein Black Hole}

Consider the five-dimensional  Einstein-Maxwell theory with action
\begin{equation}
\label{eq1}
 I=\frac{1}{16\pi G}\int_{{\cal M}}
d^{5}x\sqrt{-g}\left( R - F_{\mu\nu}F^{\mu\nu}\right) -
\frac{1}{8\pi G} \int_ {\partial {\cal M}}d^4x \sqrt{-h} K,
\end{equation}
where $R$ is scalar curvature, $F_{\mu\nu}$ is the Maxwell field
strength, $K$ is the extrinsic curvature for the boundary
${\partial {\cal M}}$ with induced metric $h$, and $G$ is the
Newtonian constant in five dimensions. The equations of motion can
be obtained by varying the action with respect to the gauge field
$A_{\mu}$  and the metric $g_{\mu\nu}$, which yields
\begin{equation}
\label{2eq4}
\partial_{\mu}\left(\sqrt{-g}F^{\mu\nu}\right)=0,
\end{equation}
\begin{equation}
\label{2eq5}
R_{\mu\nu}-\frac{1}{2}R g_{\mu\nu} = 2 \left(F_{\mu\alpha}F_{\nu}{}^{\alpha}
-\frac{1}{4}F_{\alpha\beta}F^{\alpha\beta}\right),
\end{equation}
respectively. Here $R_{\mu\nu}$ stands for Ricci tensor. These
equations of motion have a solution~\cite{Ishihara}
\begin{equation}
\label{metric}
ds^2=-f(r)dt^2+\frac{k^2(r)}{f(r)}dr^2+\frac{r^2}{4}
\left[k(r)\left(\sigma_{1}^2+\sigma_{2}^2\right)+\sigma_{3}^2\right],
\end{equation}
where $f(r)$ and $k(r)$ are given by
\begin{equation}
f(r)=\frac{(r^2-r_{+}^2)(r^2-r_{-}^2)}{r^4},\quad k(r)
=\frac{(r_{\infty}^2-r_{+}^2)(r_{\infty}^2-r_{-}^2)}{(r_{\infty}^2-r^2)^2},
\end{equation}
$r_{+},r_{-}$ and $r_{\infty}$ are three integration constants
which satisfy $r_{-}<r_{+}<r_{\infty}$, and
\begin{eqnarray}
\sigma_{1}&=&-\sin{\psi}d\theta+\cos{\psi}\sin{\theta}d\phi \nonumber \\
\sigma_{2}&=&\cos{\psi}d\theta+\sin{\psi}\sin{\theta}d\phi \nonumber \\
\sigma_{3}&=&d\psi+\cos{\theta}d\phi,
\end{eqnarray}
are Maurer-Cartan one form on $S^3$, where $0<\theta <\pi$,
$0<\phi<2\pi$, $0<\psi<4\pi$ are Euler angles. The gauge potential
is given by
\begin{equation}
\label{eq7}
 A=\pm\frac{\sqrt{3}}{2}\left(
\frac{r_{+}r_{-}}{r^2}-\frac{r_{+}r_{-}}{r_{\infty}^2}\right)dt,
\end{equation}
It is worth stressing here that this gauge potential is different
from the one in \cite{Ishihara} by  the second term on the right
hand side of (\ref{eq7}).  This difference is just a gauge
transformation and will not affect the Lagrangian and equations of
motion because there are no Chern-Simons terms in the action. With
this term, one finds that the gauge potential vanishes at
spatial infinity ($r\rightarrow r_{\infty}$). Furthermore, we will
see in section 4, this change will be important to satisfy the first
law of thermodynamics for the Kaluza-Klein black hole.

The function $k(r)$ in the solution (\ref{metric}) represents the
deformation of the black hole horizon. Clearly for $k(r)=1$
(namely, $r_{\infty} \to \infty$; in this case, the solution is
just a five-dimensional Reissner-Nordtsr\"om black hole) the
horizon is a round three-sphere $S^3$, while the horizon has
topology $S^1\hookrightarrow S^2$ when $k(r) \ne 1$. From the
solution it appears to have four singularities at $r=0$, $r_{\pm}$
and $r_{\infty}$, respectively. As shown in \cite{Ishihara},
however, the intrinsic singularity is just the one at $r=0$.
$r_{\pm}$ and $r_{\infty}$  are coordinate singularities;
$r_{\pm}$ are in fact the outer and inner horizons of the black
hole and $r\to r_{\infty}$ is a spatial infinity. To see this,
one can define a new radial coordinate $\rho$ as~\cite{Ishihara}
\begin{equation}
\rho =\rho_0 \frac{r^2}{r_{\infty}^2-r^2},\quad \rho_0^2 = k_0
\frac{r_{\infty}^2}{4},
\end{equation}
where
\begin{equation}
k_0=k(r=0)= f_\infty= f(r=r_\infty)
=\frac{(r_{\infty}^2-r_+^2)(r_{\infty}^2-r_-^2)}{r_{\infty}^4}.
\end{equation}
The new coordinate $\rho$ varies from $0$ to $\infty$
when $r$ varies from $0$ to $r_{\infty}$.
The metric (\ref{metric}) can be rewritten by $\rho$ and
$\tau=\sqrt{f_\infty}~t$ as
\begin{equation}
ds^2= - V d\tau^2 +\frac{K^2}{V} d\rho^2
+R^2 d\Omega_{S^2}^2 + W^2 \chi^2,
\label{metric_far}
\end{equation}
where $\chi=\sigma_3$, and $V, K, R$ and $W$ are functions of $\rho$
given by
\begin{eqnarray}
&& V=\frac{(\rho-\rho_+)(\rho-\rho_-)}{\rho^2},
\quad K^2 = \frac{\rho+\rho_0}{\rho},\quad R^2=\rho^2 K^2, \nonumber \\
&& W^2=\frac{r_{\infty}^2}{4}~K^{-2}
=(\rho_0+\rho_+)(\rho_0+\rho_-)~ K^{-2}.
\end{eqnarray}
Here  we have used new parameters defined by
\begin{equation}
\rho_{\pm} = \rho_0 \frac{r_{\pm}^2}{r_{\infty}^2-r_{\pm}^2}.
\end{equation}
When $\rho \rightarrow\infty$, i.e., $r \rightarrow r_{\infty}$,
the metric (\ref{metric_far}) approaches to
\begin{equation}
ds^2 = - d\tau^2 + d\rho^2
+\rho^2 d\Omega_{S^2}^2 + \frac{r_{\infty}^2}{4} \chi^2.
\label{background1}
\end{equation}
Thus we see that this space-time is locally asymptotically flat
and has a boundary $R\times S^1\hookrightarrow S^2$. This boundary
is the same as the one for the Kaluza-Klein monopole~\cite{Mann}.
In fact, when $\rho_{\pm}=0$, the black hole solution reduces to a
Kaluza-Klein monopole solution. In this sense, the solution can be
regarded as a black hole solution with a Kaluza-Klein monopole.

\section{The Mass of Kaluza-Klein Black Hole}

\subsection{The Counterterm Method for Asymptotically Locally Flat Space-times}

In \cite{Kraus}, Kraus {\it et al.} have proposed a counterterm in
a five-dimensional asymptotically flat space-time with boundary
topology $R\times S^{3}$ or $R^{2}\times S^{2} $. By taking the
variation of the action in (\ref{eq1}) plus this counterterm
action with respect to the boundary metric one can give the
boundary stress-energy tensor~\cite{BY}, and then define the
conserved charges with some Killing vector. However, for a
five-dimensional asymptotically flat solution with a fibred
boundary topology $R^{2}\hookrightarrow S^{2}$, Mann  and Stelea
have suggested a simple counterterm~\cite{Mann} (see also
\cite{radu1,radu2} for counterterms for other boundary topologies)
\begin{equation}
I_{ct}=\frac{1}{8\pi G}\int d^{4}x\sqrt{-h}\sqrt{2\mathcal{R}} \,
, \label{countertermaction}
\end{equation}
where $\mathcal{R}$ is the Ricci scalar of the induced metric on
the boundary, $h_{ij}$.  With this counterterm, the  boundary
stress-energy tensor is found to be
\begin{equation}
\label{mann} T_{ij}=\frac{1}{8\pi G}\left( K_{ij}-Kh_{ij}-\Psi(
\mathcal{R}_{ij}-\mathcal{R}h_{ij})-h_{ij}h^{kl}\Psi_{;kl}+\Psi_{;ij}
\right)\,,
\end{equation}
where $K$ is the trace of extrinsic curvature $K_{ij}$ of the
boundary, and $\Psi=\sqrt{\frac{2}{\mathcal{R}}}$. If the boundary
geometry has an isometry generated by a Killing vector $\xi ^{i}$,
then $ T_{ij}\xi ^{j}$ is divergence free, from which it follows
that the quantity
\begin{equation}
\mathcal{Q}=\int_{\Sigma }d\Sigma_{i}T^{i}{}_{j}\xi ^{j},
\label{concharge}
\end{equation}
associated with a closed surface $\Sigma $, is conserved. In
particular, if $\xi ^{i}=\partial /\partial t$ then $\mathcal{Q}$
is the conserved mass $M$. This counterterm is essentially
equivalent to the one of Kraus {\it et al.} for boundary
$S^{2}\times R^{2}$, but it is considerably simpler than that of
Kraus {\it et. al.}. Some comments on the relation between these
two counterterms have also been given in \cite{Mann}.

\subsection{Counterterm Mass of the Solution}

We calculate the mass of the Kaluza-Klein black hole by using the
counterterm method defined by (\ref{countertermaction}) and
(\ref{concharge}) in the coordinates $(\tau,\rho,
\theta,\phi,\psi)$. After some calculations, we find
\begin{eqnarray}
8\pi G ~ T^{\tau}{}_{\tau}&=&\frac{1}{\rho^2}(\frac{\rho_0}{2}+\rho_{+}
+ \rho_{-})+O(\frac{1}{\rho^3}) \, ,\nonumber \\
8\pi G ~ T^{\psi}{}_{\psi}&=&\frac{1}{2\rho^2}(2\rho_0+\rho_{+} + \rho_{-})
+O(\frac{1}{\rho^3}) \, ,\nonumber \\
8\pi G ~ T^{\theta}{}_{\theta}&=&\frac{1}{8\rho^3}\left[\rho_0^2-\rho_{-}^2
+3\rho_{+}\rho_{-} - \rho_{+}^2
+\rho_0(\rho_{+}+\rho_{-})\right]+O(\frac{1}{\rho^{7/2}})\, ,\nonumber \\
8\pi G ~ T^{\phi}{}_{\phi}&=&\frac{1}{8\rho^3}\left[\rho_0^2-\rho_{-}^2
+3\rho_{+}\rho_{-} - \rho_{+}^2
+\rho_0(\rho_{+}+\rho_{-})\right]+O(\frac{1}{\rho^{4}})\, ,\nonumber \\
8\pi G ~ T^{\psi}{}_{\phi}&=&\frac{1}{2\rho^2}(2\rho_0+\rho_{+} + \rho_{-})
\cos{\theta}+O(\frac{1}{\rho^3}) \, .
\end{eqnarray}
Considering
\begin{equation}
d\Sigma_{\tau}\sim \rho^2 \sqrt{(\rho_{0}+\rho_{+})(\rho_{0}+\rho_{-})}
\sin{\theta} d\theta\wedge d\phi \wedge d\psi\,,
\end{equation}
and using (\ref{concharge}), we obtain the mass of the
Kaluza-Klein black hole ($G=1$)
\begin{equation}
M_{CT}=2\pi
\sqrt{(\rho_0+\rho_{+})(\rho_0+\rho_{-})}~\left(\frac{\rho_0}{2}+\rho_{+}
+ \rho_{-}\right)\, .
\end{equation}
In terms of $r_{\pm}$ and $r_{\infty}$, it can be expressed as
\begin{equation}
\label{eq20}
M_{CT}=\frac{\pi\left(r_{\infty}^4-3r_{+}^2r_{-}^2+r_{\infty}^2(r_{+}^2
+r_{-}^2)\right)}
{4\sqrt{(r_{\infty}^2-r_{+}^2)(r_{\infty}^2-r_{-}^2)}}\,.
\label{countertermmass}
\end{equation}
It is interesting to note that although one can take the limit
$r_{\infty} \to \infty$ in the solution (\ref{metric}) (in this
case, the solution reduces to the Reissner-Nordstr\"om black
hole), the mass (\ref{eq20}) diverges in this limit. This is
because $r_{\infty}$ denotes the radius $\psi$ in
(\ref{background1}), and therefore one cannot take this limit in
this set of coordinates.

\subsection{Abbott-Deser Mass for Kaluza-Klein  Black Hole}

Abbott and Deser have proposed a general definition of conserved
charges for space-times with arbitrary asymptotic
behavior~\cite{AD}. These conserved charges are associated with
the isometries of the asymptotic geometry which is supposed to be
the vacuum of the system. In \cite{ HongLu} (see also \cite{Ceb}),
the authors have generalized the construction of Abbott-Deser to
the realm of gauged supergravity theory\footnote{Although the
discussion in \cite{HongLu} is focused on the asymptotically AdS
solutions of gauged supergravity, it is easy to see that their
discussion is also applicable for the general Einstein-Maxwell
theory.}, and here we will follow their notations. The metric
$g_{\mu\nu}$ of the space-times can be decomposed as
\begin{equation}
g_{\mu\nu}= \bar{g}_{\mu\nu} + h_{\mu\nu}\,,\label{hdef}
\end{equation}
where $\bar {g}_{\mu\nu}$ is the background solution.
One can define
\begin{equation}
H^{\mu\nu} = h^{\mu\nu} - \frac{1}{2} \bar {g}^{\mu\nu}\, h^{\rho}{}_{\rho}\,,
\end{equation}
\begin{equation}
K^{\mu\nu\rho\sigma} = \frac{1}{2}( \bar {g}^{\mu\sigma}\, H^{\rho\nu} +
 \bar {g}^{\rho\nu}\, H^{\mu\sigma} - \bar {g}^{\mu\rho}\, H^{\nu\sigma}
 - \bar {g}^{\nu\sigma}\, H^{\mu\rho})\,,
\end{equation}
It should be noted,  here and in what follows, all indices are
raised and lowered using the background metric $\bar
{g}_{\mu\nu}$.

Taking $\bar{\xi}^{\mu} \partial_{\mu} = \frac{\partial}{\partial
t}$ as the canonically-normalized time-like Killing vector, the
generalized Abbott-Deser mass is then given by
\begin{equation}
M_{AD}  =  \frac{1}{8\pi}\, \oint dS_i {\cal{ M}}^i\,,
\label{ADM}
\end{equation}
where
\begin{equation}
{\cal{ M}}^i= -\sqrt{-\bar {g}}\, \Big[
\bar{\xi}_{\nu}\, \bar{\nabla}_{\mu} K^{ti\nu \mu} -
K^{tj\nu i}\, \bar{\nabla}_j \bar{\xi}_{\nu}\Big]\,,\label{admass}
\end{equation}
where $dS_i$ is the area element of the spatial surface at large
radius, the index $t$ denotes the time coordinate index, Greek
indices run over all space-time directions, and Latin indices run
over the spatial directions. Eventually, one sends the radius to
infinity and integrates at infinity.

For the Kaluza-Klein black hole with squashed horizon, it is
natural to choose (\ref{background1}) as the background solution;
therefore, we can give the $\mathcal{M}^{r}$ in the expression
(\ref{admass}) for the Abbott-Deser mass,
\begin{eqnarray}
&&-\sqrt{-\bar {g}}\, \Big[ \bar{\xi}_{\nu}\, \bar{\nabla}_{\mu} K^{tr\nu \mu} -
K^{tj\nu r}\, \bar{\nabla}_j \bar{\xi}_{\nu}\Big]
=\frac{\rho\sqrt{(\rho_{0}+\rho_{+})(\rho_{0}+\rho_{-})}}{2(\rho+\rho_{0})^2
(\rho-\rho_{+}) (\rho-\rho_{-})}\nonumber \\
&&~~~~~~~~~~\times~ \Bigg[ -2\rho_{0}^2\rho_{+}\rho_{-}
+\rho^3\Big(\rho_{0}+2(\rho_{+}+\rho_{-}) \Big)+
\rho \rho_{0}\Big(2\rho_{0}^2-5\rho_{+}\rho_{-}\nonumber \\
&&~~~~~~~~~~+2\rho_{0}(\rho_{+}+\rho_{-})\Big)+\rho^2
\Big(4\rho_{0}^2-2\rho_{+}\rho_{-}+5\rho_{0}(\rho_{+}+\rho_{-})\Big)
\Bigg]~\sin{\theta} \, .
\end{eqnarray}
When $\rho\rightarrow \infty$, we have
\begin{eqnarray}
&&-\sqrt{-\bar {g}}\, \Big[ \bar{\xi}_{\nu}\, \bar{\nabla}_{\mu} K^{tr\nu \mu} -
K^{tj\nu r}\, \bar{\nabla}_j \bar{\xi}_{\nu}\Big]\nonumber \\
&&~~~~~~~=\sqrt{(\rho_0+\rho_{+})(\rho_0+\rho_{-})}~\left(\frac{\rho_0}{2}+\rho_{+}
+
\rho_{-}\right)\sin{\theta}+\mathcal{O}\left(\frac{1}{\rho}\right)
\, .
\end{eqnarray}
Substituting this result into (\ref{ADM}), and integrating at
spatial infinity,  we get the Abbott-Deser mass of the black hole
\begin{equation}
\label{eq28}
 M_{AD}=2\pi
\sqrt{(\rho_0+\rho_{+})(\rho_0+\rho_{-})}~\left(\frac{\rho_0}{2}+\rho_{+}
+ \rho_{-}\right)\, .
\end{equation}
This result agrees precisely with the mass obtained in
(\ref{countertermmass}) by using the boundary counterterm method.
Therefore, if we choose the metric (\ref{background1}) as the
background solution, we have
\begin{equation}
M_{CT}=M_{AD} \, .
\end{equation}
As we said before, in some cases if one uses a certain method
which is background dependent, to get conserved charges, different
background choice will result in different result.  For the
Kaluza-Klein black hole considered in this paper, if we use the
solution (\ref{metric_far}) with $\rho_{\pm}=0$, but keeping
$\rho_{0}$ finite as the background solution, using the
generalized Abbott-Deser method, we find that the mass of the
black hole becomes
\begin{equation}
M_{AD}'= \frac{\pi}{2}[3\rho_0 (\rho_++\rho_-)-\rho_+\rho_-],
\end{equation}
or
\begin{equation}
\label{eq31}
M_{AD}'=\frac{\pi}{8r^2_{\infty}}[3r^2_{\infty}(r_+^2+r_-^2)-7r^2_+r^2_-],
\end{equation}
in terms of $r_{\pm}$ and $r_{\infty}$. Clearly it is different
from that in (\ref{eq28}) for the choice of the solution
(\ref{background1}) as the background.
We will show that (\ref{eq28}) is more sensible because it satisfies
the first law of thermodynamics.

\section{Thermodynamics of the Kaluza-Klein Black Hole}

Although the horizon of the Kaluza-Klein black hole is deformed by
the function $k(r)$, the entropy associated with the black hole
horizon still obeys the area formula. To see this, it is very
convenient to use Wald's formula~\cite{Wald}.  It turns out that
the entropy of the black hole is
\begin{equation}
\label{eq32}
 S=4\pi^2
(\rho_{+})^{\frac{3}{2}}(\rho_0+\rho_{-})^{\frac{1}{2}}(\rho_0+\rho_{+})
=\frac{\pi^2~r_{+}^3}{2}\cdot
\frac{r_{\infty}^2-r_{-}^2}{r_{\infty}^2-r_{+}^2}.
\end{equation}
To derive the Hawking temperature of the black hole, one may use
the Euclidean method~\cite{GH}.  Through continuing the black hole
solution (\ref{metric_far}) to its Euclidean sector, to avoid the
conical singularity at the horizon, the Euclidean time coordinate
has to have a fixed period. The period is just the inverse Hawking
temperature ($1/T$). For the black hole solution
(\ref{metric_far}), it gives us
\begin{equation}
\label{eq33}
T=\frac{\rho_{+}-\rho_{-}}{4\pi
\rho_{+}^2}\sqrt{\frac{\rho_{+}}{\rho_{+}+\rho_{0}}}
=\frac{r_{+}^2-r_{-}^2}{2\pi~r_{+}^3}\frac{r_{\infty}^2}{r_{\infty}^2-r_{-}^2}
\sqrt{\frac{r_{\infty}^2-r_{+}^2}{r_{\infty}^2-r_{-}^2}} \, .
\end{equation}
Of course, one can also get the Hawking temperature by calculating
the surface gravity ($\kappa$) on the horizon via the formula,
$T=\kappa/2\pi$. We note from (\ref{eq32}) and (\ref{eq33}) that
when $r_{\infty} \to \infty$, they reduces to the entropy and
Hawking temperature of a five-dimensional Reissner-Nordstr\"om
black hole, respectively.  When $r_+=r_-$, the Hawking temperature
vanishes, which corresponds to the extremal limit of the black
hole. Interestingly enough, we note that when $r_+ \to r_{\infty}$,
the black hole entropy diverges and temperature vanishes.
Considering the fact that $r_{\infty}$ denotes the radius of the
coordinate $\psi $ in (\ref{background1}) and $r \to r_{\infty}$
represents the spatial infinity, one can easily understand the
behavior of the limit $r_+ \to r_{\infty}$: the limit means that
the Kaluza-Klein black hole has an infinite horizon radius; as a
result, the entropy diverges and temperature approaches to zero,
like a Schwarzschild black hole with infinite radius.

In the coordinates $(\tau,\rho,\theta,\phi,\psi)$, the gauge potential
on the horizon becomes
\begin{eqnarray}
\Psi&=&A_{\tau}=A_{t}\left(\frac{\partial \tau}{\partial t}\right)^{-1}
=\Phi (f_{\infty})^{-\frac{1}{2}}
=\pm \frac{\sqrt{3}}{2} \frac{ r_{-}}{r_{+}}
\sqrt{\frac{(r_{\infty}^2-r_{+}^2)}{ (r_{\infty}^2-r_{-}^2)}}\nonumber \\
&=&\pm\frac{\sqrt{3}}{2}\sqrt{\rho_+ \rho_-}
\left(\frac{\rho_0}{\rho_0+\rho_+}\right)
\left(\frac{1}{\rho_{+}}+\frac{1}{\rho_{0}}\right)\, .
\end{eqnarray}
The electrical charge (defined by
$Q=-\frac{1}{4\pi}\int_{S^3}{}^{*} F $) does not change with this
change of coordinates, so we have
\begin{equation}
Q~=~ \pm \frac{\sqrt{3}}{2} \pi \, r_{+}r_{-} = \pm 2\sqrt{3} \pi
\sqrt{\rho_{+}\rho_{-}(\rho_0+\rho_{+})(\rho_0+\rho_{-})} \, .
\end{equation}
It can be checked that if one uses the mass (\ref{eq20}), these
quantities satisfy the first law of black hole thermodynamics
\begin{equation}
\label{eq36}
dM- T~ dS -\Psi~ dQ = 0.
\end{equation}
In this formula, we have considered $r_{+}$, $r_{-}$ as variables
and $r_{\infty}$ as a constant. If we allow $r_{\infty}$ variable,
there will be an additional term $Pdr_{\infty}$ in (\ref{eq36}).
Since $r_{\infty}$ can be regarded as a deformation parameter of
the horizon, the conjugate quantity $P$ to this parameter can be
viewed as a generalized pressure. In addition, let us stress that
if one uses the mass (\ref{eq31}), the first law (\ref{eq36}) does
not hold, and that without the second term on the right hand
side of (\ref{eq7}), the first law does not hold either. One may
wonder why the first law of thermodynamics depends on the gauge
choice of the electric potential. Indeed, as we mentioned above,
the gauge choice will not affect the equations of motion and the
black hole solution. However, the first law of black hole
thermodynamics relates some conserved charges like the mass and
electric charge measured at infinity to some quantities on the
horizon like the surface gravity and entropy. The electric
potential entering into the first law should be measured relative
to infinity. In the usual case like Reissner-Nordstr\"om black
hole, the electric potential vanishes at infinity. As a result the
electric potential entering into the first law is just its
value on the horizon. In our case for the Kaluza-Klein black hole,
without the second term in (\ref{eq7}), the potential does not
vanishes at infinity ($r=r_{\infty}$), and thus in the first law
(\ref{eq36}), the electric potential should be the difference
between the horizon and infinity. Equivalently one can directly
take a gauge so that the electric potential vanishes at infinity
like (\ref{eq7}). This is the reason why we added the additional term
in  (\ref{eq7}).

\begin{figure}[htb]
\centering
\begin{minipage}[c]{.58\textwidth}
\centering
\includegraphics[width=\textwidth]{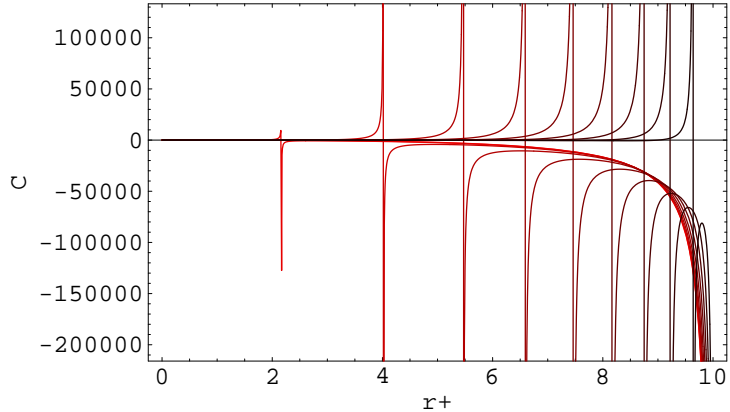}
\caption{The heat capacity $C_{Q}$ vs. $r_{+}$ of the Kaluza-Klein
black hole. Here we have set $r_{\infty}=10$. Different curves
correspond to the cases of $r_{-}=0,1,2 \cdots 8 ,9$. The color of
the curves changes from red to black when $r_{-}$ increases  from
$0$ to $9$.} \label{specialheat}
\end{minipage}
\end{figure}

\begin{figure}[htb]
\centering
\begin{minipage}[c]{.58\textwidth}
\centering
\includegraphics[width=\textwidth]{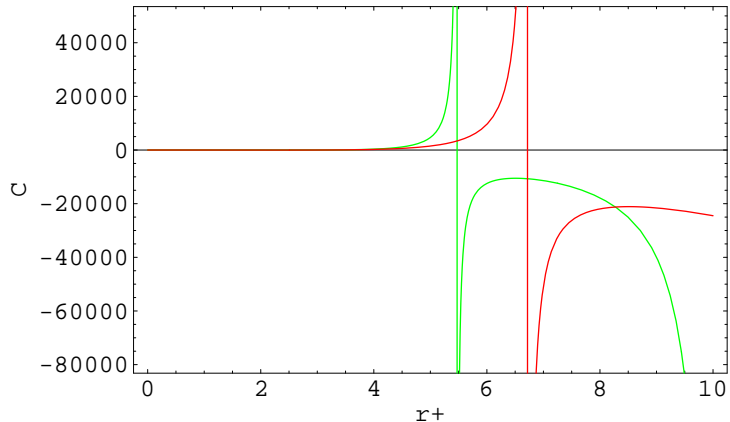}
\caption{Comparison of the heat capacity $C_{Q}$ vs. $r_{+}$ for Kaluza-Klein
black hole and Reissner-Nordstr\"om black hole. Here we have set $r_{\infty}=10$,
$r_{-}=3 $ for the Kaluza-Klein black hole and $r_{\infty}=\infty$, $r_{-}=3$
for the Reissner-Nordstr\"om black hole. The red and green curves represent
Reissner-Nordstr\"om black hole and Kaluza-Klein black hole respectively.}
\label{kkvsrn}
\end{minipage}
\end{figure}

Next let us discuss the thermodynamic stability of the black hole.
The heat capacity of the black hole for a fixed $Q$ is
\begin{equation}
C_{Q} =T\left(\frac{\partial S}{\partial T}\right)_{Q}=\frac{\pi^2
r_{+}^3}{2}
\cdot\frac{r_{\infty}^2-r_{-}^2}{r_{\infty}^2-r_{+}^2}\cdot
\frac{(r_{+}^2-r_{-}^2)(3 r_{\infty}^4-r_{+}^2r_{-}^2-r_{\infty}^2
(r_{+}^2+r_{-}^2))}
{r_{\infty}^4(5r_{-}^2-r_{+}^2)-r_{-}^2(2r_{\infty}^2-r_{+}^2)(3r_{+}^2+r_{-}^2)}
\, .
\end{equation}
Clearly this quantities goes to the one for a five-dimensional
Reissner-Nordstr\"om black hole as $r_{\infty} \to \infty$. In
order for the black hole to be stable, we must require $C_{Q}>0$. Since
\begin{equation}
3 r_{\infty}^4-r_{+}^2r_{-}^2-r_{\infty}^2(r_{+}^2+r_{-}^2)>0\, ,
\end{equation}
this means that $r_{+}$ must satisfy
\begin{equation}
r_{\infty}^4(5r_{-}^2-r_{+}^2)-r_{-}^2(2r_{\infty}^2-r_{+}^2)(3r_{+}^2+r_{-}^2)>0\,
, \label{criticalequation}
\end{equation}
i.e.
\begin{equation}
r_{+} < r_{crit}=\sqrt{\frac{1}{6r_{-}^2}\left[
r_{\infty}^4+6r_{\infty}^2r_{-}^2-r_{-}^4-(r_{\infty}^2-r_{-}^2)
\sqrt{r_{\infty}^4+14r_{\infty}^2r_{-}^2+r_{-}^4}\right]}\,
,
\end{equation}
where $r_{crit}$ corresponds to the phase transition point of
Davies~\cite{Davies}: one has $C_Q>0$ for $r_+ <r_{crit}$, $C_Q<0$
for $r_+>r_{crit}$ and the heat capacity diverges when the horizon
radius $r_+$ crosses the critical value $r_{crit}$. We plot the
heat capacity versus the horizon radius $r_{+}$ in
Fig.~\ref{specialheat} . A comparison with the
Reissner-Nordstr\"om black hole is made in Fig.~\ref{kkvsrn}. We
see from figures that when $r_+ \to r_{\infty}$, $C_Q$ goes to
minus infinity like a Schwarzschild black hole, as expected.

We can consider some limits for the heat capacity of the
Kaluza-Klein black hole.

\indent (i). Take the limit $r_{\infty}\rightarrow \infty$, then
we have
\begin{equation}
C_{Q}=\frac{3\pi^2 r_{+}^3(r_{+}^2-r_{-}^2)}{2(5r_{-}^2-r_{+}^2)},
\quad r_{crit}=\sqrt{5} ~r_{-}\,.
\end{equation}
These are just the heat capacity and critical point of a
five-dimensional Reissner-Norstr\"om black hole. In the limit, the
black hole solution reduces to a five-dimensional
Reissner-Nordstr\"om black hole.

\indent (ii). When $r_{-}\rightarrow 0$, namely the case without
charge, there are no critical points. This case corresponds to a
monotonic curve in Fig.1. One can understand this from
Eq.~(\ref{criticalequation}). In this limit, the heat
capacity becomes
\begin{equation}
C_{Q}=\frac{\pi^2 r_{+}^3 (r_{+}^2-3r_{\infty}^2)}{2(r_{\infty}^2-r_{+}^2)}
\end{equation}
and it further gives the  heat capacity of a five-dimensional
Schwarzschild black hole when $r_{\infty}\rightarrow \infty$. When
$r_+=0$, the  heat capacity vanishes. This is consistent with the
fact that there is no thermodynamic properties for the
Gross-Perry-Sorkin monopole~\cite{Mann}.

\indent (iii). When $r_{+}=r_{-}$, the  heat capacity always
vanishes. This corresponds to the case of extremal Kaluza-Klein
black holes with vanishing Hawking temperature.

\section{Conclusion and Discussion}

In this paper, we have calculated the mass of the Kaluza-Klein
black hole with squashed horizon by using the counterterm method
and generalized Abbott-Deser method. Since the generalized
Abbott-Deser method is background dependent, only when a suitable
background (\ref{background1}) is chosen, it gives the same mass
as the counterterm method.  Only this mass has been shown to
satisfy the first law of thermodynamics for the Kaluza-Klein black
hole. For example, the mass given in (\ref{eq31}), which is
obtained by choosing the solution (\ref{metric_far}) with
$\rho_{\pm}=0$ but keeping $\rho_0$ finite as the reference
background, does not satisfy the first law.  Further, we have
noticed that the electric potential entering into the first law of
the black hole thermodynamic is the potential difference between
the black hole horizon and infinity. If without the second term in
(\ref{eq7}), the potential does not vanish at infinity
($r=r_{\infty})$; this is quite different from the case of
Reissner-Nordstr\"om  black hole solution. In (\ref{eq7}) we have
taken a gauge so that the electric potential vanishes at infinity.
In addition, we have discussed some thermodynamic properties of
the Kaluza-Klein black holes, and some interesting limits have
been found.

\section*{Acknowledgments}
LMC thanks Hong-Sheng Zhang, Hao Wei, Hui Li, Da-Wei Pang and Yi
Zhang for useful discussions and kind help. This work was finished
during RGC visits the department of physics, Osaka university
through an JSPS invited fellowship, the warm hospitality extended
to him is appreciated. This work is supported by grants from NSFC,
China (No. 10325525 and No. 90403029), and a grant from the
Chinese Academy of Sciences. NO was supported in part by the
Grant-in-Aid for Scientific Research Fund of the JSPS No.
16540250.



\begin{thebibliography}{99}

\bibitem{Haw}S.~W.~Hawking,
Commun.\ Math.\ Phys.\  {\bf 25}, 152 (1972).

\bibitem{ER}R.~Emparan and H.~S.~Reall,
  Phys.\ Rev.\ Lett.\  {\bf 88}, 101101 (2002)
  [arXiv:hep-th/0110260].

\bibitem{Ishihara}
H. Ishihara and K. Matsuno, arXiv:hep-th/0510094.

\bibitem{AD}
L. F. Abbott and S. Deser, Nucl. \ Phys. \ B {\bf 195}, 76 (1982)

\bibitem{GH}
G. W. Gibbons and S. W. Hawking, Phys. \ Rev. \ D {\bf 15}, 2752 (1977)

\bibitem{BY}
J. D. Brown and J. W. York, Phys. \ Rev. \ D {\bf 47}, 1407 (1993)

\bibitem{bk}V. Balasubramanian and P. Kraus,
Commun.\ Math.\ Phys.\ {\bf 208}, 413 (1999)
[arXiv:hep-th/9902121].

\bibitem{Kraus} P.~Kraus, F.~Larsen and R.~Siebelink, Nucl.\ Phys.\ B \textbf{%
563}, 259 (1999) [arXiv:hep-th/9906127].

\bibitem{Mann}
R. B. Mann and C. Stelea, arXive:hep-th/0511180.


\bibitem{radu1}D.~Astefanesei and E.~Radu,
  Phys.\ Rev.\ D {\bf 73}, 044014 (2006)
  [arXiv:hep-th/0509144].

\bibitem{radu2}B.~Kleihaus, J.~Kunz and E.~Radu,
  arXiv:hep-th/0603119.


\bibitem{HongLu}
W. Chen, H. L\"u  and C. N. Pope, arXive:hep-th/0510081.

\bibitem{Ceb}
H.~Cebeci, O.~Sarioglu and B.~Tekin,
  Phys.\ Rev.\ D {\bf 73}, 064020 (2006)
  [arXiv:hep-th/0602117].

\bibitem{Wald}R.~M.~Wald,
  Phys.\ Rev.\ D {\bf 48}, 3427 (1993)
  [arXiv:gr-qc/9307038].

\bibitem{Davies}P.~C.~W.~Davies,
  Proc.\ Roy.\ Soc.\ Lond.\ A {\bf 353}, 499 (1977).

\end{thebibliography}
\end{document}